%Paper: hep-th/9305114
%From: melzer@max.physics.sunysb.edu (Ezer Melzer)
%Date: Sun, 23 May 1993 15:42:24 -0500 (EDT)

%
% This tex file uses harvmac
%
\input harvmac
\def\ie{{\it i.e.}}
\def\eg{{\it e.g.}}

\def\no{\noindent}
\def\o{\over}
\def\nl{\hfill\break}
\def\alp{\alpha}\def\del{\delta}

\def\sig{\sigma}

%\font\numbers=cmu10 scaled\magstep1

\def\IR{\relax{\rm I\kern-.18em R}}
\font\cmss=cmss10 \font\cmsss=cmss10 at 7pt
\def\ZZ{\relax\ifmmode\mathchoice
{\hbox{\cmss Z\kern-.4em Z}}{\hbox{\cmss Z\kern-.4em Z}}
{\lower.9pt\hbox{\cmsss Z\kern-.4em Z}}
{\lower1.2pt\hbox{\cmsss Z\kern-.4em Z}}\else{\cmss Z\kern-.4em Z}\fi}
\def\inbar{\,\vrule height1.5ex width.4pt depth0pt}
\def\IC{\relax\hbox{$\inbar\kern-.3em{\rm C}$}}
\def\IN{\relax{\rm I\kern-.18em N}}
\def\IQ{\relax\hbox{$\inbar\kern-.3em{\rm Q}$}}
\def\bQ{{\bf Q}}
\def\tbQ{\tilde{{\bf Q}}}
\def\bA{{\bf A}}
\def\bu{{\bf u}}
\def\tbu{\tilde{{\bf u}}}
\def\bm{{\bf m}}
\def\be{{\bf e}}
\def\bz{{\bf z}}

\def\sL{^{(L)}}
\def\sz{^{(0)}}
\def\so{^{(1)}}
\def\sLo{^{(L-1)}}
\def\sLt{^{(L-2)}}
\def\tF{\tilde{F}}

\def\brho{{\rho\kern-0.465em \rho}}
\def\atd{\atopwithdelims[]}
\def\atdr{\atopwithdelims()}
\def\ontopss#1#2#3#4{\raise#4ex \hbox{#1}\mkern-#3mu {#2}}

%\draft

\nref\rRog{L.J. Rogers, Proc. London Math. Soc. (ser. 1) 25 (1894) 318;
  (ser. 2) 16 (1917) 315.}
\nref\rRR{L.J. Rogers and S. Ramanujan, Proc. Camb.
 Phil. Soc. 19 (1919) 211.}
\nref\rSchur{I. Schur, S.-B. Preuss. Akad. Wiss. Phys.-Math. Kl.
 (1917) 302.}
\nref\rSchuri{I. Schur, S.-B. Preuss. Akad. Wiss. Phys.-Math. Kl.
 (1926) 488.}
\nref\rGorAnd{B. Gordon, Amer. J. Math. 83 (1961) 363; \nl
 G.E. Andrews, Proc. Nat. Sci. USA 71 (1974) 4082.}
\nref\rAndrbook{G.E. Andrews, {\it The Theory of Partitions}
  (Addison-Wesley, London, 1976).}
\nref\rBaxi{R.J. Baxter, J. Stat. Phys. 26 (1981) 427.}
\nref\rVertex{{\it Vertex Operators in Mathematics
 and Physics}, ed. J.~Lepowsky, S. Mandelstam and I.M. Singer
 (Springer, Berlin, 1985).}
\nref\rOJM{M. Okado, M. Jimbo and T. Miwa, Sugaku Expositions 2 (1989) 29.}
\nref\rKMM{R.~Kedem, B.M.~McCoy and E.~Melzer,
 Stony Brook preprint, hep-th/9304056,
 to appear in C.N.~Yang's 70th birthday Festschrift, ed. S.T. Yau.}
\nref\rBPZ{A.A. Belavin, A.M. Polyakov and A.B. Zamolodchikov,
 Nucl. Phys. B241 (1984) 333.}
\nref\rFQS{D. Friedan, Z. Qiu and S.H. Shenker, Phys. Rev. Lett.
 52 (1984) 1575.}
\nref\rGKO{P.~Goddard, A. Kent and D. Olive, Commun. Math. Phys. 103
 (1986) 105.}
\nref\rRoCa{A. Rocha-Caridi, in~\rVertex.}
%: {\it Vertex Operators in Mathematics
% and Physics}, ed. J.~Lepowsky, S. Mandelstam and I.M. Singer
% (Springer, Berlin, 1985).}
\nref\rmath{V.G. Kac, {\it  Infinite dimensional Lie algebras}, third
 edition (Cambridge University Press, 1990);\nl
 A.J. Feingold and J. Lepowsky, Adv. in Math. 29 (1978) 271; \nl
 B.L Feigin and D.B. Fuchs, Funct. Anal. Appl. 17 (1983) 241;\nl
 V.G. Kac and D.H. Peterson, Adv. in Math. 53 (1984) 125; \nl
 M. Jimbo and T. Miwa, Adv. Stud. in Pure Math. 4 (1984) 97; \nl
 V.G.~Kac and M.~Wakimoto, Adv.~in Math.~70 (1988) 156.}
\nref\rBMP{P. Bouwknegt, J. McCarthy and K. Pilch, Nucl. Phys. B352
 (1991) 139.}
\nref\rFelder{G. Felder, Nucl. Phys. B317 (1989) 215.}
\nref\rLepPrim{J.~Lepowsky and M.~Primc, {\it Structure of the
 standard modules for the affine Lie algebra $A_1^{(1)}$},
 Contemporary Mathematics, Vol.~46 (AMS, Providence, 1985).}
\nref\rLepWil{J. Lepowsky and R.L. Wilson, Proc. Nat. Acad. Sci. USA
 78 (1981) 7254; Adv. in Math. 45 (1982) 21; Invent. Math. 77 (1984) 199.}
\nref\rFNO{B.L. Feigin, T. Nakanishi and H. Ooguri, Int. J. Mod.
 Phys. A7, Suppl. 1A (1992) 217}
\nref\rBCNA{H.W.J. Bl\"ote, J.L. Cardy and M.P. Nightingale, Phys. Rev.
 Lett. 56 (1986) 742; ~I. Affleck, Phys. Rev. Lett. 56 (1986) 746.}
\nref\rCardy{J.L.~Cardy, Nucl.~Phys.~B270 (1986) 186.}
\nref\rModInv{A.~Cappelli, C.~Itzykson and J.-B.~Zuber,
 Nucl.~Phys.~B280 (1987) 445; \nl
 D.~Gepner, Nucl.~Phys.~B287 (1987) 111.}
\nref\rPI{C. Itzykson and J.-B. Zuber, Nucl. Phys. B275 (1986) 580;\nl
 D.Z. Freedman and K. Pilch, Ann. Phys. 192 (1989) 331.}
\nref\rFerFi{A.E. Ferdinand and M.E. Fisher, Phys. Rev. 185 (1969) 832.}
\nref\rKedMc{R.~Kedem and B.M.~McCoy, Stony Brook preprint,
 hep-th/9210129, J. Stat. Phys.~(in press).}
\nref\rDKMM{S. Dasmahapatra, R. Kedem, B.M. McCoy and E. Melzer,
 Stony Brook preprint, hep-th/9304150.}
\nref\rDKKMM{S. Dasmahapatra, R.~Kedem, T.R.~Klassen, B.M.~McCoy and
 E.~Melzer, Stony Brook/Rutgers preprint,
 hep-th/9303013, to appear in the proceedings of ``Yang-Baxter Equations
 in Paris'', ed. J.-M. Maillard (World Scientific).}
\nref\rRSNRT{B.~Richmond and G.~Szekeres, J.~Austral.~Math.~Soc.~(ser. A)~31
             (1981) 362;\nl
 W.~Nahm, A.~Recknagel and M.~Terhoeven, Bonn preprint, hep-th/9211034.}
\nref\rKKMMi{R.~Kedem, T.R.~Klassen, B.M.~McCoy and E.~Melzer,
 Phys. Lett. B304 (1993) 263 (hep-th/9211102).}
\nref\rTerKNS{M.~Terhoeven, Bonn preprint, hep-th/9111120;\nl
 A. Kuniba, T. Nakanishi, J. Suzuki, Harvard preprint,
 hep-th/9301018.}
\nref\rKKMMii{R.~Kedem, T.R.~Klassen, B.M.~McCoy and E.~Melzer,
 Stony Brook/Rutgers preprint, hep-th/9301046,
 Phys. Lett. B (in press).}
\nref\rBaxbook{R.J. Baxter, {\it Exactly Solved Models in Statistical
 Mechanics} (Academic Press, London, 1982).}
\nref\rFSN{O. Foda, Nijmegen preprint (1991); \nl
 K.A. Seaton and B. Nienhuis, Nucl. Phys. B384 (1992) 507.}
\nref\rSB{H. Saleur and M. Bauer, Nucl. Phys. B320 (1989) 591.}
\nref\rCardyii{J.L. Cardy, Nucl. Phys. B275 (1986) 200.}
\nref\rAndii{G.E. Andrews, Scripta Math. 28 (1970) 297.}
\nref\rABF{G.E. Andrews, R.J. Baxter and P.J. Forrester, J. Stat.
 Phys. 35 (1984) 193.}
\nref\rChKRV{P. Christe, Int. J. Mod. Phys. A6 (1991) 5271;
 ~J. Kellendonk, M. R\"osgen and R. Varnhagen, Bonn preprint,
 hep-th/9301086.}
\nref\rSlater{L.J. Slater, Proc. London Math. Soc. (ser. 2)
 54 (1951-52) 147.}
\nref\rKauf{B. Kaufman, Phys. Rev. 76 (1949) 1232.}
\nref\rADM{G. Albertini, S. Dasmahapatra and B.M. McCoy, Int. J. Mod. Phys.
 A7, Suppl. 1A (1992) 1.}
\nref\rHuse{D.A. Huse, Phys. Rev. B30 (1984) 3908.}
\nref\rFenGin{P. Fendley and P. Ginsparg, Nucl. Phys. B324 (1989) 549.}
\nref\rPasq{V. Pasquier, Nucl. Phys. B285 (1987) 162; J. Phys. A20
 (1987) L217.}
\nref\rher{T.R. Klassen and E. Melzer, Cornell/Stony Brook preprint,
 hep-th/9206\-114, Int. J. Mod. Phys. A (in press).}

\Title{\vbox{\baselineskip12pt\hbox{ITP-SB-93-29}
\hbox{hep-th/9305114} } }
{\vbox{\centerline{Fermionic Character Sums and}
\vskip 4mm
       \centerline{the Corner Transfer Matrix}}}

\centerline{Ezer Melzer~\foot{Address after Sept.~1, 1993:
 School of Physics and Astronomy, Tel-Aviv University, Tel-Aviv 69978,
 Israel.}}

\medskip \medskip
{\vbox{\baselineskip14pt{\centerline{{\it Institute for Theoretical Physics}}
\centerline{{\it State University of New York}}
\centerline{{\it Stony Brook,  NY 11794-3840}} } }

%\centerline{email: melzer@max.physics.sunysb.edu}

\vskip 4mm

\centerline{\bf Abstract}
\medskip

We present a ``natural finitization'' of the fermionic $q$-series
(certain generalizations of the Rogers-Ramanujan sums) which were
recently conjectured to be equal to Virasoro characters of the unitary
minimal conformal field theory (CFT) ${\cal M}(p,p+1)$. Within the
quasi-particle interpretation of the fermionic $q$-series this
finitization amounts to introducing an ultraviolet cutoff,
which -- contrary to a lattice spacing -- does not modify the linear
dispersion relation. The resulting polynomials are conjectured
(proven, for $p$=3,4) to be equal to corner transfer matrix (CTM) sums
which arise in the computation of order parameters in regime III of
the $r$=$p$+1 RSOS
model of Andrews, Baxter, and Forrester.  Following Schur's proof of the
Rogers-Ramanujan identities, these authors have shown that the
infinite-lattice limit of the CTM sums gives what later became known
as the Rocha-Caridi formula for the Virasoro characters. Thus we
provide a proof of the fermionic $q$-series representation for the
Virasoro characters for $p$=4 (the case $p$=3 is ``trivial''), in
addition to extending the remarkable connection between CFT and
off-critical RSOS models.  We also discuss finitizations of the CFT
modular-invariant partition functions.

\Date{}

\vfill\eject

\newsec{Prelude}

This paper is concerned with generalizations of the following identities:
\eqn\rsr{ \eqalign{
 \sum_{m=0}^{\infty}{q^{m(m+a)}\o (q)_m}~ & =~
  \prod_{n=1}^{\infty}~{1\o (1-q^{5n-4+a})(1-q^{5n-1-a})} \cr
 &=~ \lim_{L\to\infty} \sum_{\sig_i\in\{0,1\},~\sig_i \sig_{i+1}=0
 \atop \sig_1=a, ~\sig_{L+1}=0}  q^{\sum_{j=1}^{L-1} j\sig_{j+1}} \cr
 &= ~{1\o (q)_{\infty}}
 \sum_{k\in \ZZ}\left(q^{k(10k+1+2a)}-q^{(2k+1)(5k+2-a)}\right)~ \cr}}
for $a=0,1$,
where\foot{Throughout the paper, $\ZZ$, $\IR$, and $\IC$ denote the sets
of all integers, real numbers, and complex numbers, respectively.
Also $\ZZ_N=\ZZ/(N\ZZ)$ is the cyclic (additive) group of $N$ elements.}
\eqn\qsub{(q)_0=1~~,~~~~~~~  (q)_{m}=\prod_{n=1}^m (1-q^n)~~~~~~{\rm for}~~
  m=1,2,3,\ldots.}

Though equal, the four sides of \rsr\
represent for us different objects. In order to distinguish
between them in the discussion below, we will refer to
(generalizations of) the infinite
sums on the first, second, and third lines of \rsr\ as
{\it fermionic}, {\it corner-transfer-matrix} (CTM),
and {\it bosonic}, respectively. (Since \rsr\ contains only
a single infinite product expression it is not necessary
to give it a name; if pressed, we will
call it a bosonic product.) In order to avoid the
introduction of unnecessary notation which might obscure the
general picture at an early stage, we defer the
presentation of the specific generalizations of \rsr\ we are
interested in to the main body of the paper; cf.~eqs.~(3.20) and
(3.22) for the fermionic sums, (3.12) for the CTM sums,
and (2.2) for the bosonic ones.

The first line of \rsr, stating the equality (fermionic sum = product),
is due to Rogers~\rRog\rRR, Schur~\rSchur, and Ramanujan~\rRR, and
is usually referred to as the Rogers-Ramanujan(-Schur) identities.
The equality  (product = CTM sum)
has a combinatorial interpretation in terms of certain restricted
partitions of integers~\rSchur -\rGorAnd~(see Corollaries 7.6--7.7
in~\rAndrbook). A direct proof of (CTM sum = fermionic sum)
is given in~\rBaxi, and finally the bosonic side is found
in~\rRog -\rSchur.

\newsec{Introduction}

The relevance of the celebrated identities \rsr\
and their generalizations
to two-dimensional physics, as well as string theory, has been
recognized in the last two decades; see \eg~\rVertex~and the
reviews~\rOJM\rKMM.
The ``physics connection'' goes through (at least) two independent
routes, which we will call the critical and the off-critical.
Describing these connections will motivate the names -- completely
arbitrary, up to this point -- which were associated
above with the sums in \rsr.

\subsec{The critical connection}

It turns out that many $q$-series of the type
\rsr\ are characters of irreducible
highest-weight representations of infinite-dimensional algebras,
called {\it chiral} or {\it vertex operator} algebras,
which serve as symmetry algebras of two-dimensional conformal field
theories (CFTs).
In particluar, the two $q$-series in \rsr\ (with $a$=0 and 1)
are the Virasoro characters corresponding to the two primary fields
(of conformal dimensions $\Delta=-{1\o 5}$ and 0, respectively)
of the minimal model ${\cal M}(2,5)$.
This model belongs to the minimal series~\rBPZ\
of models ${\cal M}(p,p')$, labeled by two coprime positive integers
$p'>p\geq 2$. The central charge and highest-weights of the
corresponding irreducible representations ${\cal V}^{(p,p')}_{~r,s}$
of the Virasoro algebra are\foot{Whenever the superscript $p'$ is suppressed
in formulas below, it will be understood as being equal to $p+1$,
corresponding to the unitary series~\rFQS\rGKO ~${\cal M}(p,p+1)$.}
{}~$c^{(p,p')}=1-{6(p'-p)^2\o pp'}$~ and
\eqn\Delrs{\Delta_{~r,s}^{(p,p')}~=~{(rp'-sp)^2-(p'-p)^2\o 4pp'}
 ~~~~~~~(r=1,\ldots,p-1,~~s=1,\ldots,p'-1).}
The characaters of these representations
are~\rRoCa
\eqn\Virchi{\chi^{(p,p')}_{~r,s}(q) \equiv
  q^{-\Delta_{~r,s}^{(p,p')}}{\rm Tr}_{{\cal V}_{~r,s}^{(p,p')}}~q^{L_0}=
 {1\o (q)_{\infty}}
  \sum_{k\in \ZZ}\bigl(q^{k(kpp'+rp'-sp)}-q^{(kp+r)(kp'+s)}\bigr),}
where the overall normalization is chosen such that
$\chi^{(p,p')}_{~r,s}(q)=1+\sum_{n=1}^\infty a_n q^n$,
with the $a_n$ non-negative integers which depend on
{}~$p,p',r,s$.
Note the symmetry of the ``conformal grid''
$\{(r,s)~|~r\in\{1,\ldots,p-1\},~s\in\{1,\ldots,p'-1\}\}$,
\eqn\gridsym{(r,s)\leftrightarrow (p-r,p'-s):
{}~~~\Delta_{~r,s}^{(p,p')}=\Delta_{p-r,p'-s}^{(p,p')}~~
 \Rightarrow ~~\chi_{~r,s}^{(p,p')}(q)=\chi_{p-r,p'-s}^{(p,p')}(q)~.}
The bosonic sum in  \rsr\ is
precisely ~$\chi_{1,2-a}^{(2,5)}(q)=\chi_{1,3+a}^{(2,5)}(q)$, as given
by the rhs of \Virchi.

\medskip

We remark that ``purely mathematical'' derivations (see
\eg~\rGKO -\rmath)
of characters of highest-weight
representations of chiral algebras, as well as their
computation using free-field resolutions (\rBMP~and
references therein) based on the work of~\rFelder,
usually lead to bosonic expressions for the characters.
The signature of these expressions are factors
of $(q)^{-1}_\infty$
which is  the character of a (freely generated)
bosonic Fock space.
[Note, however, that when the free-field
realization of the chiral algebra involves Fermi fields, as
is the case for
superconformal algebras, the resulting character formulas also
include factors of $\prod_{n=1}^\infty(1+q^{n-\epsilon})$.
This ``fermionic product'', to be contrasted with
$(q)^{-1}_\infty=\prod_{n=1}^\infty(1-q^n)^{-1}$ or the product
in \rsr, is the character of
a fermionic Fock space, with $\epsilon=0,{1\o 2}$ corresponding
to periodic or anti-periodic boundary conditions, respectively,
on the free massless Fermi field. For this reason it is perhaps more
appropriate to call the sum on the third line of \rsr\ a
{\it free-field} sum.]

However, direct Lie-theoretic
derivations of {\it fermionic} sum representations for
characters which are branching functions of affine Lie algebras
are known in some cases~\rLepPrim, based on the construction of
$Z$-algebras~\rLepWil.
Also, cf.~\rFNO~for an analysis which leads to product expressions for the
characters of ${\cal M}(2,p')$, the products being the ones
in the Gordon-Andrews identities~\rGorAnd~(which include the
Rogers-Ramanujan-Schur identities as a special case).

\medskip

The Hilbert space ${\cal H}$ of any conformal field theory is a direct sum
$\oplus_{i,\bar{i}} ({\cal V}_i \otimes {\cal V}_{\bar{i}})$
of products of irreducible highest-weight representations of two
commuting copies of the Virasoro algebra,
or some extended chiral algebra,
governing the ``right-moving'' (holomorphic) and ``left-moving''
(antiholomorphic) sectors. Once the multiplicities
$N_{i\bar{i}}$ in the direct sum are specified, all the remaining
information about the spectrum of the conformal field
theory ({\it i.e.}~the conformal dimensions of all the fields, which
are the eigenvalues of $L_0$ and $\bar{L}_0$, cf.~\Virchi) is
encoded in the characters. The partition function of
the conformal field theory on a torus of modulus $\tau$
is then written as
\eqn\ptfc{ Z_{\rm CFT}(q,\bar{q}) =
 |q|^{-c/12}~{\rm Tr}_{\cal H}~q^{L_0}~\bar{q}^{\bar{L}_0} =
 |q|^{-c/12}~ \sum_{i,\bar{i}} N_{i\bar{i}}~
 q^{\Delta_i}\chi_i(q)~\bar{q}^{\Delta_{\bar{i}}}\chi_{\bar{i}}(\bar{q})~,}
where $q$=$e^{2\pi i\tau}$ and $\bar{q}$=$q^*$=$e^{-2\pi i\tau^*}$.
The prefactor $|q|^{-c/12}$ accounts for the Casimir effect~\rBCNA~and
ensures modular invariance of $Z_{\rm CFT}$ for appropriately
chosen multiplicities $N_{i\bar{i}}$ (see~\rCardy\rModInv~for the case of
the minimal models ${\cal M}(p,p')$).

This rather abstract description of the CFT spectrum can be made more
concrete. One way, still within the framework of continuum quantum
field theory, is to construct $Z_{{\rm CFT}}$ as an (appropriately
regularized) euclidean path integral on the torus, see \eg~\rPI; this
of course requires knowledge of
the measure, namely an action for the CFT.
Alternatively, one can consider realizations of the given
CFT as appropriate scaling limits of certain critical two-dimensional
models of (classical) statistical mechanics, or gapless one-dimensional
(quantum) spin chains. In the first case, the CFT partition function
is obtained from the classical partition sum on a doubly-periodic $L
\times L'$ lattice in the limit $L,L'\to\infty$ with $L'/L$ fixed.
Then the variable $q$ in \ptfc\ is $e^{\alp L'/L}$, where
$\alp\in \IC$ depends on the anisotropy. In the second case $Z_{{\rm
CFT}}$ is obtained by considering the partition function of the
$L$-site spin chain at temperature $T$ in the limit
$L,T^{-1}\to\infty$ with $LT$ fixed. Now $q=e^{-2\pi v/LT}$,
where $v\in \IR$ is the fermi velocity characterizing the massless
dispersion relation of (all) the excitations. (In both cases one has
to factor out a certain ``bulk contribution'' in order to obtain
$Z_{{\rm CFT}}$; see \eg~sect.~3 of \rDKKMM.) Clearly
it is very difficult to fully derive
$Z_{{\rm CFT}}$ from the definition of particular lattice/chain models.
With the exception of the Ising model~\rFerFi, results in this
direction are rather limited and often incomplete.

In~\rKedMc \rDKMM~the spectrum of the gapless three-state Potts spin
chain was analyzed in the conformal scaling limit. This analysis,
in which the eigenvalues of
the hamiltonian are computed from
solutions of Bethe equations, leads to a description of the
spectrum in terms of excitations which were called {\it
quasi-particles}.  These quasi-particles obey a fermionic exclusion
rule in momentum space, and moreover the
momentum ranges are
subject to particular restrictions which depend on the number of
quasi-particles in a given state. From this description of the
spectrum in the scaling limit, fermionic expressions for the
characters of the CFT -- of ($\ZZ_4$) $\ZZ_3$ parafermions in the
case of the (anti-) ferromagnetic three-state Potts
chain -- were obtained~\rKedMc
\rDKMM. (In the case of $\ZZ_4$ parafermions the fermionic
expressions are the ones found earlier by Lepowsky and
Primc~\rLepPrim~using Lie-theoretic methods.)
These results suggested
generalizations\foot{Important clues for obtaining these
generalizations come from the connection between the analysis of the
leading $q$$\to$1 behavior of the fermionic
sums~\rRSNRT -\rKKMMii~and
thermodynamic Bethe Ansatz computations.
This connection, which involves
sum rules for the Rogers dilogarithm function, is beyond
the scope of the present article.} which led to the
discovery/conjecture of new fermionic sum representations for many
classes of CFT characters~\rKMM\rKKMMi -\rKKMMii,
all of them allowing an interpretation of the
CFT spectrum in terms of fermionic quasi-particles.

\subsec{The off-critical connection}

CTM sums have been encountered in computations of order parameters
in off-critical exactly solvable models using Baxter's corner transfer matrix
technique~\rBaxbook. In particular, the sum on the second line of \rsr\
appears in the analysis of regime I of the generalized hard-hexagon
model~\rBaxi~(in addition, as noted in~\rBaxi,
the ratio of the products
with $a=0,1$ emerged already in the elliptic parametrization of the
Boltzmann weights which define the model!). The fact that CTM
sums are equal to characters of chiral algebras
seems {\it a priori} very mysterious.
Note in particular that the variable $q$ in the CTM context is a
temperature-like parameter, which measures
the deviation from criticality ($q_c$=1).
Hence it appears to have nothing to do
with any of the three ``critical meanings'' of $q$ mentioned in
subsect.~2.1. Nevertheless, the connection has been recently
elucidated in~\rFSN~based on the work of~\rSB. The key idea
(see~\rFSN~for the details) is to
relate the order parameters, which are one-point functions
on the plane with fixed boundary conditions at ``infinity'', to
the partition function at criticality on a cylinder with fixed
boundary conditions on its rims.
CFT predicts~\rCardyii~that the partition function
in such a geometry is
a linear combination of characters, rather than the toroidal
bilinear form~\ptfc.

An interesting feature of the CTM sums is that they provide a
natural (in the CTM context) {\it finitization}\foot{We use this awkward
word in contradistiction with `truncation', to make sure that
the procedure we talk about is not confused with that of
eliminating from
the $q$-series powers which are bigger than some threshold.}
 of the $q$-series into an
infinite set of polynomials, which we call {\it CTM polynomials}. For the
particular case of \rsr, they are defined simply by removing the lim sign
from the second line of that equation:
\eqn\CLa{C^{(L)}_a(q) ~=~
 \sum_{\sig_2,\ldots,\sig_L\in\{0,1\}
  \atop \sig_1=a,~\sig_i \sig_{i+1}=0,~\sig_{L+1}=0}
  q^{\sum_{j=1}^{L-1} j\sig_{j+1}}~~~~~~~~~~~~(a=0,1)}
for ~$L=1,2,\ldots$, with
$C^{(0)}_a(q)=\del_{a,0}$.
(Here $C\sL_a(q)$ is $F(a)$ of
eq.~(44) of~\rBaxi, with $m$=$L$, so that $L$ is the size of
the edge of the corner
on which the CTM acts.)

It is natural to ask whether there exist
``natural'' finitizations of the fermionic
and bosonic sums in \rsr, which
coincide with the $C\sL_a$.
The answer is positive: For $a=0,1$~ let~\rAndii
\eqn\FLa{F\sL_a(q) ~=~ \sum_{m\in \ZZ} q^{m(m+a)}{L-m-a \atd m}_q~,}
and~\rSchur
\eqn\BLa{\eqalign{ B\sL_a(q) &=
 \sum_{\ell\in \ZZ} (-1)^\ell~q^{\ell(5\ell-1-2a)/2}
     {L \atd [{1\o 2}(L+5\ell-a)]}_q\cr
 =\sum_{k\in \ZZ}~& \Biggl( q^{k(10k+1+2a)}{L \atd [{L-a\o 2}]-5k}_q
  -q^{(2k+1)(5k+2-a)} {L \atd [{L+a\o 2}]-5k-2}_q \Biggr)~,\cr}}
where $[x]$ denotes the integer part of $x$.
Both expressions here involve
$q$-binomial coefficients, which are defined (for $m,n\in \ZZ$) by
\eqn\qbin{ {n \atopwithdelims[] m}_q ~=~ {n \atd n-m}_q~=~
 \cases{ ~{(q)_n \o (q)_m (q)_{n-m}}
  ~~~~~~~~& if ~~$0\leq m \leq n$ \cr  ~0 & otherwise~.\cr} }
The fact that $F^{(\infty)}_a(q)$ and $B^{(\infty)}_a(q)$ are
the fermionic and bosonic sums in \rsr, respectively,
follows from the properties
\eqn\infqbin{ \lim_{n\to \infty}{n \atd m}~=~{1\o (q)_m}~~~,
 ~~~~~~~~~ \lim_{n,m\to \infty}{n+m \atd m}~=~{1\o (q)_\infty}~~.}
The finitized version of \rsr\ (without a finitized product) is
\eqn\FeqCeqB{F\sL_a(q) ~=~ C\sL_a(q)~=~ B\sL_a(q)~~~~~~~~~
  {\rm for}~~~a=0,1,~~~L=0,1,2,\ldots.}

\medskip

The proof of \FeqCeqB, following~\rSchur\rSchuri\rBaxi\rAndii,
will serve as a warm-up for the more involved
discussion in the next section. The idea is to compare the recursion
relations (in $L$) and initial conditions which uniquely specify
the $C\sL_a$, $F\sL_a$, and $B\sL_a$.

\medskip

Consider first the $C\sL_a$. From the definition \CLa\ we have
for $L\geq 3$
\eqn\CLarec{\eqalign{ C^{(L)}_a ~&=~
 \sum_{\sig_2,\ldots,\sig_{L-1}\in\{0,1\}
  \atop \sig_1=a,~\sig_i \sig_{i+1}=0}
  q^{\sum_{j=1}^{L-2} j\sig_{j+1}}~
 \sum_{\sig_L\in\{0,1\} \atop \sig_{L-1} \sig_L=0=\sig_{L+1}}
  q^{(L-1)\sig_L}~\cr
  &=~(1+q^{L-1}) ~C^{(L-2)}_a~+~q^{L-2}~C^{(L-3)}_a~, \cr}}
where the first (second) term on the last line arises from
the summation on the first line when restricted to $\sig_{L-1}=0~(1)$.
The recursion relation \CLarec, the same for both $a=0,1$,
together with the initial conditions
\eqn\CLain{\eqalign{
  C^{(0)}_0 = 1~,~~~~~~C^{(1)}_0 &= 1~,~~~~~~C^{(2)}_0=1+q\cr
  C^{(0)}_1 = 0~,~~~~~~C^{(1)}_1 &= 1~,~~~~~~C^{(2)}_1=1~~~,\cr}}
are sufficient to uniquely specify all the $C^{(L)}_a$.

A derivation of a recursion relation for the $F\sL_a$
requires the use of (one of) the
following recursions  for the $q$-binomial coefficients:
\eqn\qbinrec{ {n \atd m}_q ~=~{n-1 \atd m}_q + q^{n-m}{n-1 \atd m-1}_q
 ~=~ {n-1 \atd m-1}_q + q^{m}{n-1 \atd m}_q ~~.}
We see that for $L\geq 2$
\eqn\FLarec{\eqalign{ F\sL_a ~&=~ \sum_{m\in \ZZ} q^{m(m+a)}
  \left({L-1-m-a \atd m}_q~+q^{L-2m-a}{L-1-m-a \atd m-1}_q \right)\cr
   &=~ F^{(L-1)}_a+ q^{L-1} F^{(L-2)}_a ~,\cr}}
where the second line is obtained after changing the summation variable
in the second sum on the first line from $m$ to $m'=m-1$.
It follows that \FLarec\ together with the initial conditions,
which are readily obtained from the definition \FLa,
\eqn\FLain{
  F^{(0)}_0 = 1~,~~~F^{(1)}_0= 1~~;~~~~~~~~~
  F^{(0)}_1 = 0~,~~~F^{(1)}_1= 1~,}
fully determine all the  $F\sL_a$. Now note that
by iterating \FLarec\ once (replacing the term $F^{(L-1)}_a$ on the rhs
by the rhs of \FLarec\ with $L\to L-1$),
one finds that the $F\sL_a$ actually satisfy the recursion relation
\CLarec\ for the $C\sL_a$, and since the initial conditions
\FLain\ and \CLain\
coincide
we conclude that the first equality in
\FeqCeqB\ holds. A corollary of this result is the equality
of the fermionic and CTM sums in \rsr\ (note that Baxter's
proof~\rBaxi~of this equality is obtained along different lines).

Regarding the $B\sL_a$,
we state (omitting the details, which are
slightly more involved, cf.~\rSchur\rAndii) that the recursion relations
\eqn\BLarec{ B\sL_a ~=~ B\sLo_a + q^{L-1} B\sLt_a~~~~~~~~~
  {\rm for}~~~a=0,1,~~~L=2,3,\ldots}
are obtained from the definition \BLa, using \qbinrec.
These recursions are the same as \FLarec, and also
the initial conditions for the $B\sL_a$ can be seen to
coincide with \FLain.  Hence the
equality of the $B\sL_a$ and the $F\sL_a$ in \FeqCeqB\ follows.

\medskip

\newsec{(Finitized) characters of unitary minimal models}

Up to this point we reviewed known material.
We now generalize the discussion above
to the case of the Virasoro characters
\Virchi\ with $p'$=$p$+1 (which will henceforth be suppressed
in formulas below). We start by introducing some extensive

\subsec{Notation}

Denote by $C_n$ and $I_n=2-C_n$ the Cartan and incidence matrices,
respectively, of the simple Lie algebra $A_n$, \ie
\eqn\Imat{ (I_n)_{ab} ~=~ \delta_{a,b+1} + \delta_{a,b-1} ~~~~
   ~~~~{\rm for}~~~ a,b=1,\ldots,n.}
Also let $\be_a$ denotes  the $n$-dimensional
unit vector in the $a$-direction
(\ie ~$({\bf e}_a)_b=\del_{ab}$), and set ${\bf e}_a$=0 for
$a \not\in \{1,\ldots,n\}$.
For $L$ a non-negative integer,
{}~$\bA,\bu\in \ZZ^n$,  and $\bQ \in (\ZZ_2)^n$  such that
{}~$\bQ I_n+\bu +L\be_1 \in 2\ZZ^n$, we define
\eqn\FnL{ F_n\sL{{\bf Q}\atopwithdelims[]{\bf A}}({\bf u}|q) ~=~
 \sum_{{\bf m}\in 2\ZZ^n+{\bf Q}}
  q^{{1\o 4}{\bf m} C_n {\bf m}^t -{1\o 2}{\bf A}\cdot{\bf m}}
  \prod_{a=1}^n ~
  { {1\o 2}({\bf m}I_n+{\bf u}+L\be_1)_a \atopwithdelims[] m_a}_q~~,}
where ~${\bf A}\cdot {\bf m}=\sum_{a=1}^n A_a m_a$.

Now let us fix ~$n=p-2\geq 1$, and set
\eqn\AuQrs{\eqalign{
 {\bf Q}_{r,s} ~&=~ (s-1)\brho+({\bf e}_{r-1} +{\bf e}_{r-3}+\ldots)
  + ({\bf e}_{p+1-s} +{\bf e}_{p+3-s}+\ldots)\cr
   \tbQ_{r,s} ~&=~ ({\bf e}_{p-1-r} +{\bf e}_{p-3-r}+\ldots)
  + ({\bf e}_{p-1-s} +{\bf e}_{p-3-s}+\ldots)\cr
  \bu_{r,s}~&=~\be_r+\be_{p-s}~,~~~~~~
  \tbu_{r,s}~=~\be_{p-r}+\be_{p-s}~,~~~~~~\bA_s~=~\be_{p-s}~,\cr}}
where ~$\brho={\bf e}_1+\ldots+{\bf e}_n$.
The following relations are valid for all
$r=1,2,\ldots p-1$ and  $s=1,2,\ldots,p$~
(recall that $\bQ\in (\ZZ_2)^n$, and note that $\bA_1=\bA_p=0$):
\eqn\AuQrela{
 \bQ_{r,s} I_n+\bu_{r,s}+(s-r-1)\be_1 \in 2\ZZ^n~,~~~~
  \tbQ_{r,s} I_n+\tbu_{r,s} +(s-r)\be_1 \in 2\ZZ^n}
\eqn\AuQrelb{\eqalign{
 (\tbQ_{r,1},\bA_1,\tbu_{r,1})~&=~ (\bQ_{p-r,p},\bA_p,\bu_{p-r,p})~~\cr
 (\tbQ_{r,p},\bA_p,\tbu_{r,p})~&=~ (\bQ_{p-r,1},\bA_1,\bu_{p-r,1})\cr}}
\eqn\AuQrelc{\eqalign{
 (\tbQ_{1,s},\bA_s,\tbu_{1,s})~&=~ (\bQ_{1,s},\bA_s,\bu_{1,s}-\be_1)~~\cr
(\tbQ_{p-1,s},\bA_s,\tbu_{p-1,s})~&=~(\bQ_{p-1,s},\bA_s,\bu_{p-1,s}+\be_1)
  ~~\cr.}}

Next we define the objects of most interest to us:
\eqn\FprsL{ \eqalign{
 F_{p;r,s}\sL &(q) =  q^{-{(s-r)(s-r-1)/4}} \cr
 &\times \Biggl\{
 {F_{p-2}\sL{\bQ_{r,s}\atd \bA_s}(\bu_{r,s}|q)
   ~~~~~~~~~~~~~~~~~~~~~~\hbox{if~~$L\not\equiv s-r$~(mod~2)}  \atop
 F_{p-2}\sL{\bQ_{p-r,p+1-s}\atd \bA_{p+1-s}}(\bu_{p-r,p+1-s}|q)
   ~~~~~~\hbox{if~~$L\equiv s-r$~(mod~2),}} \cr
 \tF_{p;r,s}\sL &(q) = q^{-{(s-r)(s-r-1)/4}} \cr
 &\times \Biggl\{ {F_{p-2}\sL{\tbQ_{r,s}\atd \bA_s}(\tbu_{r,s}|q)
  ~~~~~~~~~~~~~~~~~~~~~~\hbox{if~~$L\equiv s-r$~(mod~2)} \atop
 F_{p-2}\sL{\tbQ_{p-r,p+1-s}\atd \bA_{p+1-s}}(\tbu_{p-r,p+1-s}|q)
   ~~~~~~\hbox{if~~$L\not\equiv s-r$~(mod~2).} }  \cr } }
The relations \AuQrela\ ensure that these definitions
are consistent, namely that the upper entries of all the $q$-binomials
on the rhs (cf.~\FnL) are integers.
Furthermore, from the definition it is manifest that
\eqn\Fprsymm{
  F_{p;r,s}\sL(q) ~=~ F_{p;p-r,p+1-s}\sL(q)~,~~~~~~~~
  \tF_{p;r,s}\sL(q) ~=~ \tF_{p;p-r,p+1-s}\sL(q)~,}
and the relations \AuQrelb--\AuQrelc\ imply
\eqn\FtF{
   \tF_{p;r,1}\sL(q) ~=~ F_{p;r,1}\sL(q)~,~~~~
  \tF_{p;1,s}\sL(q) ~=~ \cases{
     F_{p;1,s}^{(L+1)}(q)~~~ &~~~if~~$L\equiv s$ (mod 2) \cr
     F_{p;1,s}^{(L-1)}(q)~~~ &~~~if~~$L\not\equiv s$ (mod 2)~. \cr}}

\medskip

We move on to a different set of definitions, essentially
borrowed from the work of
Andrews, Baxter, and Forrester~\rABF~on the RSOS model in regime III. For
{}~$a,b,c\in\{1,2,\ldots,p\}$ with  $|b-c|=1$, and $L$ a non-negative
integer such that ~$L\equiv a-b$ (mod 2), let
\eqn\preCpL{C\sL_p(a,b,c;q) ~=~
 \sum_{h_2,\ldots,h_L\in\{1,\ldots,p\}
  \atop (h_1,h_{L+1},h_{L+2})=(a,b,c),~|h_i-h_{i+1}|=1}
  q^{\sum_{j=1}^{L} j|h_{j+2}-h_j|/4}~.}
(This sum is precisely $X_L(a,b,c;q)$, eq.~(1.5.11) of~\rABF~with
$r$=$p$+1.) The following symmetry property is obvious
(just change summation variables from ~$h_i$~ to ~$p+1-h_i$):
\eqn\presym{ C\sL_p(a,b,c;q) ~=~ C\sL_p(p+1-a,p+1-b,p+1-c;q)~.}
Next define for ~$r=1,2,\ldots,p-1$~ and ~$s=1,2,\ldots,p$
\eqn\CpLrs{
 C_{p;r,s}\sL(q) =q^{-(s-r)(s-r-1)/4} \cdot \cases{
 C_{p}\sL(s,r,r+1;q)
   &if~~$L\equiv s-r$ (mod 2) \cr
 C_{p}\sL(s,r+1,r;q)
   ~~~&if~~$L\not\equiv s-r$ (mod 2). \cr}}
The power of $q$ here is such that the $C_{p;r,s}\sL(q)$ are polynomials
in $q$, which satisfy, as follows from \presym,
\eqn\Cprsymm{  C_{p;r,s}\sL(q) ~=~ C_{p;p-r,p+1-s}\sL(q)~.}

{}From the analysis in subsect.~2.3 of~\rABF~one can infer
that the $C_{p;r,s}\sL(q)$ are uniquely determined from the
recursion relation
\eqn\CpLrec{ C_{p;r,s}\sL(q) = C_{p;r,s}^{(L-1)}(q)
   + \cases{ q^{(L+s-r)/2} C_{p;r-1,s}^{(L-1)}(q)
                &~~if~~$L\equiv s-r$ (mod 2) \cr
             q^{(L-s+r+1)/2} C_{p;r+1,s}^{(L-1)}(q)
                &~~if~~$L\not\equiv s-r$ (mod 2) \cr} }
(where it is understood that $C_{p;r,s}\sL(q)$=0 for
$r\not\in\{1,\ldots,p-1\}$), and the initial conditions
\eqn\CpLin{ C_{p;r,s}^{(0)}(q) ~=~\del_{r,s}+\del_{r+1,s} ~.}

Using the above definitions, Theorem 2.3.1 of~\rABF~can be restated as
\eqn\CeqBpL{ C_{p;r,s}\sL(q) ~=~ B_{p;r,s}\sL(q)~,}
where the $B_{p;r,s}\sL(q)$ are defined by
\eqn\BpLrs{\eqalign{ B_{p;r,s}\sL(q)~ =~
  \sum_{k\in \ZZ} \Biggl(   q^{k(kp(p+1)+r(p+1)-sp)}
                    &{L \atd [{L+s-r\o 2}]-k(p+1) }_q \cr
          -~q^{(kp+r)(k(p+1)+s)}
            &{L \atd [{L-s-r\o 2}]-k(p+1) }_q \Biggr)~.\cr}}
The proof of Theorem 2.3.1 of~\rABF, as given there, demonstrates
that the $B_{p;r,s}\sL(q)$ satisfy \CpLrec--\CpLin,
which implies the identity \CeqBpL.

\subsec{Main claim}

We are now in a position to state our
\eqn\conj{\eqalign{  &{\underline {\rm Conjecture}}:~~~~~~~
   \tF_{p;r,s}\sL(q) ~=~ F_{p;r,s}\sL(q) ~=~ B_{p;r,s}\sL(q) ~=~
      C_{p;r,s}\sL(q) \cr
   &{\rm for~all}~~~~L=0,1,2,\ldots,~~p=3,4,5,\ldots,~~
   r=1,2,\ldots,p-1,~~s=1,2,\ldots,p.\cr} }
(Of course the third equality is just \CeqBpL, which was proved
in~\rABF.) Before describing the evidence we have in support
of \conj, cf.~subsect.~3.4, let us discuss its
implications for

\subsec{Infinite $L$}

When $L$ becomes infinite, we see from \infqbin\ that
$F_n\sL{\bQ \atd \bA}(\bu |q)$ reduces to
\eqn\Sn{ S_n{\bQ \atd \bA}(\bu|q) ~=
 \sum_{ { {\bf m}\in 2\ZZ^n+{\bf Q} \atop m_1\geq 0}  }
  q^{{1\o 4}{\bf m} C_n {\bf m}^t -{1\o 2}{\bf A}\cdot{\bf m}}
  ~{1\o (q)_{m_1}}~\prod_{a=2}^n ~
  { {1\o 2}({\bf m}I_n+{\bf u})_a \atopwithdelims[] m_a}_q~~,}
whereas $B_{p;r,s}\sL(q)$ of \BpLrs\ becomes
the Virasoro character
$\chi_{r,s}^{(p)}(q)$, as given by the rhs of \Virchi\ with $p'$=$p$+1.
Therefore, using \FprsL, the second equality in \conj\
is equivalent when $L\to\infty$ to
\eqn\chieqS{ \eqalign{
 \chi_{r,s}^{(p)}(q)~&=~q^{-(s-r)(s-r-1)/4}
 S_{p-2}{\bQ_{r,s}\atd \bA_s}(\bu_{r,s}|q) \cr
    &=~q^{-(s-r)(s-r-1)/4}
 S_{p-2}{\bQ_{p-r,p+1-s}\atd \bA_{p+1-s}}(\bu_{p-r,p+1-s}|q)~. \cr}}

[The two expressions on the right here
correspond to the limits taken
with $L$ even or odd. Their equality
is immediate for the ``corners'' of the conformal grid,
$(r,s)$=(1,1) or (1,$p$) and their mirror partners ($p-1$,$p$) and
($p-1$,1); the reason is that the difference between
$(\bQ_{1,1},\bA_1,\bu_{1,1})$=(0,0,$\be_1$)
and $(\bQ_{p-1,p},\bA_p,\bu_{p-1,p})$=(0,0,0), and between
$(\bQ_{1,p},\bA_p,\bu_{1,p})$=($\be_{p-2}+\be_{p-4}+\ldots$,0,$\be_1$)
and $(\bQ_{p-1,1},\bA_1,\bu_{p-1,1})$=($\be_{p-2}+\be_{p-4}+\ldots$,0,0),
is just in the first component $u_1$ of $\bu$ on which the
$S_{p-2}{\bQ \atd \bA}(\bu |q)$ do not depend at all.
However, for other $(r,s)$ the second
equality in \chieqS\ appears to be quite nontrivial.]

\medskip

Eq.~\chieqS\ was conjectured in~\rKKMMii, based
on the analysis of the spectrum of the ferromagnetic
three-state Potts chain~\rDKMM~which
corresponds to the case $p$=5. It was verified to hold as an equality of
power series to high order in $q$ in many cases.
As another consistency check, pertaining to the infinitely high
powers in the $q$-series, the leading $q\to 1$ behavior of
the $S_{p-2}{\bQ \atd \bA}(\bu|q)$ was shown in~\rKKMMii~to
agree with the one obtained from the modular properties~\rCardy~of
the  characters $\chi^{(p)}(q)$ as determined from \Virchi, namely
\eqn\qtoi{ S_{p-2}{\bQ \atd \bA}(\bu|q)~\sim~ \tilde{q}^{-c^{(p)}/24}
 ~~~~~~~~~{\rm as}~~~~~q\to 1^-~,}
where ~$\tilde{q}$=$e^{-2\pi i/\tau}$~ for ~$q$=$e^{2\pi i\tau}$.

\medskip

Similarly to \chieqS, equality of the $\tF\sL_{p;r,s}$ and the
$B\sL_{p;r,s}$ in
\conj\ leads, when $L\to \infty$, to
\eqn\chieqSt{ \eqalign{
 \chi_{r,s}^{(p)}(q)~&=~q^{-(s-r)(s-r-1)/4}
 S_{p-2}{\tbQ_{r,s}\atd \bA_s}(\tbu_{r,s}|q) \cr
    &=~q^{-(s-r)(s-r-1)/4}
 S_{p-2}{\tbQ_{p-r,p+1-s}\atd \bA_{p+1-s}}(\tbu_{p-r,p+1-s}|q)~. \cr}}
For all $(r,s)$ along the ``boundary'' of the conformal grid,
\ie~for $r\in\{1,p-1\}$ and/or $s\in\{1,p\}$, this equation simply
reduces to \chieqS, due to the relations \AuQrelb--\AuQrelc.
For all other pairs $(r,s)$, constituting the ``interior'' of the
conformal grid, eq.~\chieqSt\ gives new fermionic sum representations
for the characters $\chi_{r,s}^{(p)}$~ (for $p$=5, $s$=3 they were found
in~\rDKMM.)

\medskip

To summarize, eqs.~\chieqS\ and \chieqSt\ are conjectured
to provide fermionic sums of the form \Sn\
for the Virasoro characters $\chi_{r,s}^{(p)}$, one for each
of the characters at the ``corners'' of the conformal grid, two
different (looking) sums
for each one of all the other characters along the ``boundary'',
and four different fermionic sums for
characters in the ``interior'' of the conformal grid.
(Of course one should in fact restrict attention to half of
the conformal grid, due to the symmetry \gridsym.)
Bosonic expressions for the characters are given in \Virchi,
and CTM forms are obtained by taking the $L\to\infty$ limit
of \CpLrs. Infinite product expressions for some of the characters
of each model ${\cal M}(p,p+1)$ can be found in~\rChKRV.

The main motivation for the present work is to try and
prove \chieqS\ and \chieqSt\ by considering the (finitized)
fermionic sums $F\sL_{p;r,s}$, $\tF\sL_{p;r,s}$ of
eq.~\FprsL\ at

\subsec{Finite $L$}

Using Mathematica, we have checked that for many choices of the
parameters $p,r,s,L$ the  $F\sL_{p;r,s}(q)$
and $\tF\sL_{p;r,s}(q)$ are equal and satisfy the same recursion
relations and initial conditions as the ones \CpLrec--\CpLin\
satisfied by the $C\sL_{p;r,s}(q)$. These successful ``experiments''
encourage an attempt to prove these recursion relations analytically.
At present, we have accomplished this task only for $p=3,4$, as we now
describe.

\bigskip \no
$\bullet$~~$p=3$ (Ising): ~This case is rather simple, and in fact an
analysis of the finitized sums is not really necessary for proving
\conj\ at $L=\infty$. The $F^{(\infty)}_{3;r,s}(q)$  and
$\tF^{(\infty)}_{3;r,s}(q)$  can be expressed as (linear
combinations of) certain infinite products -- see finitized
versions below -- through simple combinatorial considerations.
The corresponding identities (see \eg~item (2) on
Slater's list~\rSlater, known as Euler's identity),
and the equality of the infinite
products and the bosonic sums $B^{(\infty)}_{3;r,s}(q)$
(cf.~\rRoCa), are well known. Nevertheless, we think that
the forthcoming analysis is illuminating.

There are three characters in this case, which can be taken to
be $\chi^{(3)}_{1,1}$, $\chi^{(3)}_{1,2}$, and $\chi^{(3)}_{2,1}$.
According to \FprsL--\FtF\
we have to consider only four sums. Explicitly, they are
\eqn\Fth{\eqalign{
 F\sL_{3;1,1}(q)=\sum_{m\in 2\ZZ}
                 q^{{m^2\o 2}}{[{L+1\o 2}] \atd m}_q~~&,~~~
 F\sL_{3;1,2}(q)=\sum_{m\in 2\ZZ +L+1}
                 q^{{m(m-1)\o 2}}{ [{L+2\o 2}] \atd m}_q \cr
 F\sL_{3;2,1}(q)=\sum_{m\in 2\ZZ +1}
                 q^{{m^2-1\o 2}}{[{L+1\o 2}]\atd m}_q~~&,~~~
 \tF\sL_{3;1,2}(q)=\sum_{m\in 2\ZZ +L}
                 q^{{m(m-1)\o 2}}{ [{L+2\o 2}] \atd m}_q~~.\cr}}
Using \qbinrec, followed by a change of summation variable
$(m-1) \to m$ in one of the two resulting sums, one obtains
the recursion relations
\eqn\Fthrec{\eqalign{
   F\sL_{3;1,1} = F\sLo_{3;1,1}+\cases{0 &\cr
         q^{{L+1\o 2}} F\sLo_{3;2,1} &\cr} ~&,~~~~
   F\sL_{3;1,2} = \tF\sLo_{3;1,2}+\cases{
         q^{{L\o 2}} F\sLo_{3;1,2}  &\cr
         0                          &\cr} \cr
   F\sL_{3;2,1} = F\sLo_{3;2,1}+\cases{0 &\cr
         q^{{L-1\o 2}} F\sLo_{3;1,1}&\cr} ~&,~~~~
 \tF\sL_{3;1,2} = F\sLo_{3;1,2}+\cases{
    q^{{L\o 2}} \tF\sLo_{3;1,2}  &\cr
    0                            &\cr} \cr} }
where the upper (lower) cases apply when $L$ is even (odd).
Hence it is sufficient to supplement them by the initial
conditions
\eqn\Fthin{
  F^{(0)}_{3;1,1}~=~F^{(0)}_{3;1,2}~=~\tF^{(0)}_{3;1,2}~=~1~,~~~~~~~
  ~F^{(0)}_{3;2,1}~=~0~,}
which are obtained from \Fth.
Comparing with \CpLrec--\CpLin\ for $p$=3, we conclude
that the unique solution to \Fth--\Fthin\  is
\eqn\FtheqC{
  F\sL_{3;1,1}=C\sL_{3;1,1}~,~~~~~
  F\sL_{3;1,2}=\tF\sL_{3;1,2}=C\sL_{3;1,2}~,~~~~~
  F\sL_{3;2,1}=C\sL_{3;2,1}~,}
which completes the proof of \conj\ for this case.

\medskip

For completeness, let us write down finite product formulas
for (linear combinations of) the $F\sL_{3;r,s}$, which can be deduced
from \Fthrec--\Fthin:
\eqn\Fthprod{\eqalign{
  F\sL_{3;1,1}(q)+q^{1/2} F\sL_{3;2,1}(q) ~& =
       \prod_{n=1}^{[{1\o 2}(L+1)]} (1+q^{n-1/2}) \cr
  F\sL_{3;1,1}(q)-q^{1/2}F\sL_{3;2,1}(q) ~& =
       \prod_{n=1}^{[{1\o 2}(L+1)]} (1-q^{n-1/2}) \cr
  F\sL_{3;1,2}(q)~=~\tF\sL_{3;1,2}(q)~ & =~~
    \prod_{n=1}^{[{1\o 2}L]} (1+q^{n})~~. \cr}}
These expressions -- in the $L$$\to$$\infty$ limit -- correspond to the
familiar construction of the Ising CFT characters in terms of
a single {\it free} massless chiral (right-moving, say) Majorana fermion,
with certain projections on sectors of even or odd number of particles
and periodic or anti-periodic boundary conditions. The finitization
of the products here simply corresponds in this language to
putting a cutoff on the allowed single-particle momenta
(or energies). In fact, {\it all} the fermionic sums -- both
infinite and finitized -- have fermionic quasi-particle
interpretations which generalize the above picture,
cf.~\rKedMc\rDKMM\rKKMMi\rKKMMii, but generically no
``fermionic product'' expressions of the type \Fthprod\ seem
to be available. It is apparently the free nature of the Majorana fermion
underlying the Ising model~\rKauf~which is responsible for this
further simplification.

As a final comment on the $p$=3 case, note
that \Fthprod\ immediately gives
\eqn\Fthqone{
  F\sL_{3;1,1}(1)=F\sL_{3;2,1}(1)  = 2^{[{1\o 2}(L-1)]}~~,~~~~~~
  F\sL_{3;1,2}(1)=\tF\sL_{3;1,2}(1) = 2^{[{1\o 2}L]}~.}
This can of course be seen also directly from \Fth, as the
$q$-binomial coefficient reduces to the ordinary  binomial
coefficient when $q$=1, as well as from the definition of the
CTM polynomials $C\sL_{3;r,s}(q)$ (see sect.~4).

\bigskip \no
$\bullet$~~$p=4$: ~Using the symmetries \Fprsymm\
and the first relation in \FtF,
the number of sums which must be considered in this case is reduced
to nine. We take them to be $F\sL_{4;r,s}$ with
$r$=1,2,3 and $s$=1,2, and $\tF\sL_{4;r,2}$ with $r$=1,2,3,
as defined by \FprsL. After some tedious elementary algebra
we arrive at the following recursion relations
(we use \qbinrec,
twice for cases with $r$=2, followed by changes of summation
variables $(m_1-1)\to m_1$~ or ~$(m_2-1)\to m_2$ whenever necessary):
\eqn\Fforeca{\eqalign{
   F\sL_{4;1,1} &= \cases{F\sLo_{4;1,1} &\cr
       F\sLt_{4;1,1}+q^{{L+1\o 2}} F\sLo_{4;2,1} &\cr} ,~~~
   F\sL_{4;3,1} = \cases{F\sLt_{4;3,1}+q^{{L-2\o 2}} F\sLo_{4;2,1} &\cr
       F\sLo_{4;3,1} &\cr} \cr
   F\sL_{4;1,2} &= \cases{F\sLt_{4;1,2}+q^{{L\o 2}}F\sLo_{4;2,2} &\cr
      \tF\sLo_{4;1,2} &\cr} ~~,~~~
   F\sL_{4;3,2} = \cases{\tF\sLo_{4;3,2} &\cr
       F\sLt_{4;3,2}+q^{{L-1\o 2}}F\sLo_{4;2,2} &\cr} \cr
 \tF\sL_{4;1,2} &= \cases{\tF\sLt_{4;1,2}+q^{{L\o 2}}\tF\sLo_{4;2,2} &\cr
       F\sLo_{4;1,2} &\cr} ~~,~~~
 \tF\sL_{4;3,2} = \cases{F\sLo_{4;3,2} &\cr
       \tF\sLt_{4;3,2}+q^{{L-1\o 2}}\tF\sLo_{4;2,2} &\cr} \cr}}
\eqn\Fforecb{\eqalign{
   F\sL_{4;2,1}&=\cases{F\sLt_{4;2,1}+q^{{L-2\o 2}}F\sLt_{4;1,1}
                                     +q^{{L+2\o 2}}F\sLt_{4;3,1} &\cr
                        F\sLt_{4;2,1}+q^{{L-1\o 2}}F\sLt_{4;1,1}
                                     +q^{{L+1\o 2}}F\sLt_{4;3,1}&\cr}\cr
   F\sL_{4;2,2}&=\cases{F\sLt_{4;2,2}+q^{{L\o 2}}\tF\sLt_{4;1,2}
                                     +q^{{L\o 2}}\tF\sLt_{4;3,2} &\cr
                        F\sLt_{4;2,2}+q^{{L-1\o 2}}\tF\sLt_{4;1,2}
                                     +q^{{L+1\o 2}}\tF\sLt_{4;3,2}&\cr}\cr
 \tF\sL_{4;2,2}&=\cases{\tF\sLt_{4;2,2}+q^{{L\o 2}}F\sLt_{4;1,2}
                                +q^{{L\o 2}}F\sLt_{4;3,2} &\cr
                        \tF\sLt_{4;2,2}+q^{{L-1\o 2}}F\sLt_{4;1,2}
                                +q^{{L+1\o 2}}F\sLt_{4;3,2}~~,&\cr}\cr}}
where the upper (lower) cases apply when $L$ is even (odd).
In fact, invoking the second relation in \FtF\ one can reexpress
all the $F\sLt_{4;r,s}$ and $\tF\sLt_{4;r,s}$ in \Fforeca\ in terms
of $F\sLo_{4;r,s}$ and $\tF\sLo_{4;r,s}$.

To fully (and uniquely)
determine all the $F\sL_{4;r,s}$ and $\tF\sL_{4;r,s}$,
it suffices to specify the following initial conditions, obtained
directly from \FprsL:
\eqn\Ffoin{\eqalign{
 F\sz_{4;1,1}=F\sz_{4;1,2}=\tF\sz_{4;1,2}=F\sz_{4;2,2}=\tF\sz_{4;2,2}&=1\cr
 F\sz_{4;2,1}=F\sz_{4;3,1}=F\sz_{4;3,2}=\tF\sz_{4;3,2}&=0\cr
 F\so_{4;1,1}=F\so_{4;1,2}=\tF\so_{4;1,2}=F\so_{4;2,2}=\tF\so_{4;2,2}&=1\cr
 F\so_{4;2,1}=F\so_{4;3,2}=\tF\so_{4;3,2}=1~,~~~~F\so_{4;3,1}&=0~.\cr}}

\medskip
Now using \CpLrec\ at $p$=4 (iterated, when necessary) and \CpLin,
we verify that
\eqn\FfoeqC{\eqalign{
  F\sL_{4;1,1}=C\sL_{4;1,1}~,
   ~~F\sL_{4;1,2} &= \tF\sL_{4;1,2}=C\sL_{4;1,2}~,~~
       F\sL_{4;2,1}=C\sL_{4;2,1}~, \cr
  F\sL_{4;2,2}=\tF\sL_{4;2,2}=C\sL_{4;2,2}~,~~
       F\sL_{4;3,1} &=C\sL_{4;3,1}~,~~
          F\sL_{4;3,2}=\tF\sL_{4;3,2}=C\sL_{4;3,2} \cr}}
is a -- and therefore {\it the} -- solution of \Fforeca--\Ffoin, which
proves \conj\ for $p$=4.

\bigskip \no
$\bullet$~ $p>4$:~ For the general case we are unable
to report complete analytic results. Deriving recursion relations
for the $F\sL_{p;r,s}$ and $\tF\sL_{p;r,s}$, which ``close'' on
these polynomials with fixed $s$ and
lower $L$, are not so easily obtained.
As examples of what we were able to show for arbitrary $p$, we list
\eqn\Fprec{\eqalign{
 F\sL_{p;1,s} &= \tF\sLo_{p;1,s}+\cases{
     q^{(L-s+2)/2} F\sLo_{p;2,s} ~~&~~if~~$L\equiv s$~(mod~2)\cr
     0                      &~~if~~$L\not\equiv s$~(mod~2)\cr}\cr
 \tF\sL_{p;1,s} &= F\sLo_{p;1,s}+\cases{
     q^{(L-s+2)/2} \tF\sLo_{p;2,s} ~~&~~if~~$L\equiv s$~(mod~2)\cr
     0                      &~~if~~$L\not\equiv s$~(mod~2)\cr}\cr
 F\sL_{p;2,s} &= F\sLt_{p;2,s}+q^{(L+s-3)/2} \tF\sLt_{p;1,s}
            +q^{(L-s+3)/2} F\sLo_{p;3,s}~~~~{\rm if}~~
                         L\not\equiv s~({\rm mod}~2) \cr
 \tF\sL_{p;2,s} &= \tF\sLt_{p;2,s}+q^{(L+s-3)/2} F\sLt_{p;1,s}
            +q^{(L-s+3)/2} \tF\sLo_{p;3,s}~~~~{\rm if}~~
                         L\not\equiv s~({\rm mod}~2), \cr}}
which can be seen to be consistent with \CpLrec\ if \conj\ holds,
but is certainly not sufficient to prove the latter.

\newsec{Discussion}

The coincidence \conj\
of finitized fermionic sum representations for Virasoro
characters and finite corner-transfer-matrix sums is quite
remarkable.
The former have a ``natural interpretation'' in terms
of the spectrum of gapless spin chains, while the latter
arise in computations of order parameters in off-critical
two-dimensional systems.
Let us make some further comments on the
two different types of objects which have been claimed
(in some cases proven) to be equal.

As was demonstrated in~\rKedMc\rDKMM~(cf.~also subsec.~2.1),
fermionic character sums arise in the analysis of those
energy levels of a gapless spin chain hamiltonian
which scale as the inverse size of the system when it
becomes infinite.
The analysis, which is based on the study of Bethe equations
obtained from functional equations for the (diagonal-to-diagonal)
transfer matrix of a corresponding critical
two-dimensional lattice model,
provides an interpretation of the fermionic sums in terms
of massless quasi-particles whose momenta obey certain restrictions
in addition to a fermionic exclusion rule. In particular,
as discussed in detail in~\rKKMMii, the fermionic sums of the
form \Sn\ for characters of the unitary Virasoro minimal model
${\cal M}(p,p+1)$,
are partition functions of such a ``gas'' of massless fermionic
quasi-particles. The single-particle
momenta of these quasi-particles
are quantized in spacing of $2\pi/L$,
where $L$ is the size of the system which serves as an infrared
cutoff. These momenta are also restricted in a way (different for different
characters) which depends on the number $m_a$ of quasi-particles
of type ~$a$=1,2,$\ldots$,$n$~ in the state. The feature common to all
fermionic sums \Sn\ for a given model ${\cal M}(p,p+1)$
(where $n$=$p$$-$2),
is that for finite $m_a$ the single-particle momenta of the
$a$=1 quasi-particles can take values in a semi-infinite range
(which becomes the positive or negative half of the real axis
as $L$$\to$$\infty$,
corresponding to either right- or left-moving quasi-particles),
whereas the momenta of all
other quasi-particles are restricted
to a finite range (which shrinks to 0 when the size of the system
becomes infinite).

Finitizing the fermionic sums, namely considering the $L$-dependent
polynomials $F\sL_{p;r,s}(q)$ of
\FprsL\ instead of the $q$-series \chieqS, has the effect
of further restricting the momenta of the $a$=1 quasi-particles
to lie in a finite range, whose upper limit depends on $L$.
This procedure can be thought
of as introducing an {\it ultraviolet} cutoff, which is of order $L^0$=1
(\ie~of order $L$ in the quantization unit $2\pi/L$).
This is not to be confused with the ``original'' ultraviolet
cutoff -- the distance between neighboring sites along
the chain -- which is present in the problem of
studying the spectrum of the spin chain. It is important to
note that the effect of the built-in
chain cutoff is to restrict the
momentum to some (periodic) Brillouin zone, in which
the dependence of the single-particle energies on the momentum is
nontrivial. On the other hand,
the ultraviolet cutoff implied by the finitization
of the fermionic character sums has the appealing property of not
changing the linear dispersion relation of the quasi-particle
excitations which survive the (conformal) continuum limit.

Now the CTM polynomials $C\sL_{p;r,s}(q)$, eq.~\CpLrs,
are obtained from the sums \preCpL\ which arise in the
analysis~\rABF~of the $r$=$p$+1 RSOS model (in regime III).
In this context $L$ is the length of the
corner edge on which the corner transfer matrix acts, namely
$L$ is half the diagonal of the big square on which
the model is defined, this big square being tilted by 45$^0$
with respect to the orientation of the square plaquettes of the
lattice. (Hence, if we envisage the spectrum
of the corner transfer matrix as built out of some (fictitious)
``CTM excitations'', then $L$ serves as an infrared cutoff for
them.) It is interesting to count the number of states
which are summed over to give the CTM polynomials, namely to
evaluate these polynomials at $q$=1. They are
simply obtained from \preCpL\ as
\eqn\CpLqone{ C\sL_{p;r,s}(1)={
   ((I_p)^L)_{s,r}  ~~~~~~\hbox{if~~$L\equiv s-r$~(mod~2)}
  \atopwithdelims\{\}
   ((I_p)^L)_{s,r+1}  ~~~\hbox{if~~$L\not\equiv s-r$~(mod~2)} }
    = ((I_p)^L)_{s,r}+((I_p)^L)_{s,r+1}~,}
where $I_p$ is the incidence matrix of $A_p$ (cf.~\Imat), which
enforces the RSOS restrictions $h_i\in\{1,\ldots,p\}$,
$|h_i-h_{i+1}|$=1 in \preCpL. (The second equality in \CpLqone\
is valid since ~$((I_p)^L)_{r,s}$=0~ if ~$(L+s+r)$ is odd.)
Our claim \conj\ then implies,
using the definitions \FprsL, \FnL, and \BpLrs,
the following infinite set of
identities for the (ordinary) binomial coefficients
${n\atdr m}={n!\o m!(n-m)!}$~
(with ${n\atdr m}$=0~ if ~$n<0$~ or
{}~$m\not\in\{0,1,\ldots,n\}$):
\eqn\idens{\eqalign{
 \sum_{\bQ_{r,s}}~ &\prod_{a=1}^{p-2}
   {{1\o 2}(\bm I_{p-2}+\bu_{r,s}+L\be_1)_a \atdr m_a} \cr
 &= \sum_{\tbQ_{p-r,p+1-s}}~ \prod_{a=1}^{p-2}
   {{1\o 2}(\bm I_{p-2}+\tbu_{p-r,p+1-s}+L\be_1)_a \atdr m_a} \cr
 &=~\sum_{k\in \ZZ} \Biggl( {L \atdr {L+s-r-1\o 2}-k(p+1)}
                           -{L \atdr {L-s-r-1\o 2}-k(p+1)}\Biggr) \cr
     &=~((I_p)^L)_{s,r+1}  ~~~~~~~~~~~~~~~
      ~~~~~~~~~~~~{\rm for}~~~(L+s+r)~~{\rm odd} \cr
 & {} \cr
 \sum_{\tbQ_{r,s}}~ &\prod_{a=1}^{p-2}
   {{1\o 2}(\bm I_{p-2}+\tbu_{r,s}+L\be_1)_a \atdr m_a} \cr
 &= \sum_{\bQ_{p-r,p+1-s}}~ \prod_{a=1}^{p-2}
   {{1\o 2}(\bm I_{p-2}+\bu_{p-r,p+1-s}+L\be_1)_a \atdr m_a} \cr
 &=~\sum_{k\in \ZZ} \Biggl( {L \atdr {L+s-r\o 2}-k(p+1)}
                           -{L \atdr {L-s-r\o 2}-k(p+1)} \Biggr) \cr
     &=~((I_p)^L)_{s,r} ~~~~~~~~~~~~~~~
       ~~~~~~~~~~~~~~~{\rm for}~~~(L+s+r)~~{\rm even}, \cr} }
where $\sum_\bQ$ indicates summation over $\bm\in 2\ZZ^{p-2}+\bQ$,
{}~$L$ is a non-negative integer, ~$p\geq 3$ an integer,
{}~$r$=1,$\ldots,p-1$, ~$s$=1,$\ldots,p$,
and the various vectors are defined in \AuQrs. We note that when $p$=3
the elements of the matrix $(I_3)^L$ can be simplified, as in \Fthqone,
and that our analysis in subsect.~3.4 also provides a proof of
\idens\ for $p$=4.

\medskip

Identities similar to \idens\ for the case of $p$=5 were encountered
in~\rADM~in the analysis of completeness of the solutions to the Bethe
equations for the three-state Potts spin chain. In particular,
one of the results proven there (eq.~(5.12) of~\rADM)
reads in our notation
\eqn\idADM{ F^{(L)}_{5;1,1}(1)+F^{(L)}_{5;1,5}(1)=
 \sum_{{m_1+m_3\in 2\ZZ \atop m_2\in 2\ZZ}}
  {{1\o 2}(L+m_2)\atdr m_1}{{1\o 2}(m_1+m_3)\atdr m_2}
  {{1\o 2}m_2\atdr m_3}= 3^{{L\o 2}-1}, }
for $L$ even.
Our conjectured identities \idens\ can be seen to be consistent
with this, as $((I_5)^L)_{1,1} +((I_5)^L)_{5,1}=
 {1\o 2}(3^{{L\o 2}-1}+1) +{1\o 2}(3^{{L\o 2}-1}-1) =3^{{L\o 2}-1}$
{}~for $L$ even.

It should be noted, though, that when analyzing completeness
of the Bethe-equation solutions, {\it all} excitations (involving both
left- and right-moving quasi-particles) are included. Each finitized
character sum, on the other hand, describes just a certain sector
of one chiral half of the theory. Recalling the factorization
\ptfc\ of the CFT partition function into characters, we are faced
with an intriguing question of whether there is some ``natural''
finitization of the partition function as a whole, where the
building blocks are the finitized characters. In the case
$p$=3 we can construct the following
two finitized partition functions, corresponding to
the Ising model and the theory of a free Majorana fermion
with antiperiodic boundary conditions
(we take $q\in \IR$ here):
\eqn\finZth{ \eqalign{
  Z_{\rm Ising}\sL(q) &= q^{-1/24}
   \sum_{s=1}^3 q^{2\Delta_{1,s}^{(3)}}F\sLo_{3;1,s}(q) ~F\sL_{3;1,s}(q)\cr
  Z_{\rm Maj}\sL(q) &= q^{-1/24}
 \bigl(  F\sLo_{3;1,1}(q) + q^{1/2} F\sLo_{3;2,1}(q) \bigr)
 \bigl(  F\sL_{3;1,1}(q) + q^{1/2} F\sL_{3;2,1}(q) \bigr)~, \cr}}
where $\Delta_{1,s}^{(3)}=0,{1\o 16}$, and ${1\o 2}$~ for ~$s$=1,2, and 3,
respectively. One reason why we regard these specific finitizations to be
natural is the existence -- for $L$ even -- of the following neat product
forms, which can be obtained from \Fthprod:
\eqn\finZthpr{\eqalign{
  Z_{\rm Ising}\sL(q)
   =~ {q^{-{1\o 24}}\o 2} \Bigl\{ \prod_{n=-{L-1\o2}}^{{L-1\o 2}}(1+q^{|n|})
     + &\prod_{n=-{L-1\o 2}}^{{L-1\o 2}}(1-q^{|n|})
     + q^{{1\o 8}} \prod_{n=-{L\o 2}}^{{L\o 2}}(1-q^{|n|}) \Bigr\} \cr
  Z_{\rm Maj}\sL(q)
   =  q^{-{1\o 24}} &\prod_{n=-{L-1\o 2}}^{{L-1\o 2}}(1+q^{|n|})~~, \cr }}
where the product variable $n$ is incremented in steps of one.
Another nice property, valid for either parity of $L$, is
\eqn\Zthqone{   Z_{\rm Ising}\sL(1) ~=~
  Z_{\rm Maj}\sL(1) ~=~ 2^L~,}
which is the total number of states of an Ising/Majorana spin chain
of $L$ sites.

We emphasize that from the point of view of the RSOS model {\it per se},
the ``completeness rule'' \Zthqone\ is somewhat surprising: recall
that the $F\sL_{3;r,s}$ were in fact constructed
to be equal to the CTM polynomials \CpLrs\ which arise in the RSOS model
of~\rABF, at $r$=4 in this case. The origin of this rule lies in the fact
that for $r$=4 the
system is equivalent~\rHuse~to (two decoupled copies of the)
Ising model, and so the
heights $h_i\in\{1,2,3\}$, which are restricted by $|h_i-h_{i+1}|$=1,
cf.~\preCpL, can be
traded in for the {\it un}restricted {\it two}-valued Ising spins.

There is another case where a similar phenomenon occurs, namely
at $r$=6. Here an orbifold construction~\rFenGin~leads to
the $D_4$ RSOS model of Pasquier~\rPasq, which is equivalent
at criticality to
two decoupled copies of
the three-state Potts model describing unrestricted
three-valued ``spins''.
We are therefore tempted to write down the obvious generalization
of $Z\sL_{\rm Ising}$ to the $\ZZ_3$ case, obtained
by finitizing the partition function of the three-state Potts
CFT~\rCardy:
\eqn\finZfi{ \eqalign{
  Z_{\rm 3sP}\sL(q) = q^{-1/15}
  \Bigl\{  &\bigl( F\sLo_{5;1,1}(q) +q^3 F\sLo_{5;4,1}(q) \bigr)
            \bigl( F\sL_{5;1,1}(q) +q^3 F\sL_{5;4,1}(q) \bigr)\cr
  +~q^{4/5} &\bigl( F\sLo_{5;2,1}(q) +q F\sLo_{5;3,1}(q) \bigr)
            \bigl( F\sL_{5;2,1}(q) +q F\sL_{5;3,1}(q) \bigr) \cr
  +~2 &\bigl( q^{2/15} F\sLo_{5;2,3}(q)~F\sL_{5;2,3}(q)
  ~+~q^{4/3}   F\sLo_{5;1,3}(q)~F\sL_{5;1,3}(q)\bigl) \Bigr\}~.\cr}}
Amusingly, using \idens, we find for all $L$
\eqn\Zfiqone{ Z_{\rm 3sP}\sL(1) ~=~
   \bigl\{ 3^{L-1}+2\cdot 3^{L-1}\bigr\} ~=~3^L~,}
with each sector of definite $\ZZ_3$ charge contributing equally.
On the other hand, the finitized version of the partition function
of the tetracritical Ising CFT~\rCardy~is
\eqn\finZtet{   Z_{\rm tetra}\sL(q) ~=~ q^{-1/15} ~\sum_{r=1}^2
 ~\sum_{s=1}^5 q^{2\Delta_{r,s}^{(5)}}F\sLo_{5;r,s}(q) ~F\sL_{5;r,s}(q)~,}
where the dimensions $\Delta_{r,s}^{(5)}$ are given by \Delrs,
and we find that
\eqn\Ztetqone{ Z_{\rm tetra}\sL(1) ~=~3^L+1~.}

\medskip

The above observations can be generalized further to the whole
$ADE$-series~\rModInv~of modular-invariant partition functions
of the unitary minimal models ${\cal M}(p,p+1)$. Namely,
consider the corresponding finitized partition functions
\eqn\finptfc{ Z\sL_{p(X)}(q) ~=~
 q^{-c^{(p)}/12}~ \sum_{r,\bar{r}=1}^{p-1}
   ~\sum_{s,\bar{s}=1}^{p} N_{r,s;\bar{r},\bar{s}}^{p(X)}~
 q^{\Delta_{r,s}^{(p)}+\Delta_{\bar{r},\bar{s}}^{(p)}}
 ~ F\sLo_{p;r,s}(q)~F\sL_{p;\bar{r},\bar{s}}(q)~.}
Here the multiplicities $N_{r,s;\bar{r},\bar{s}}^{p(X)}$,
corresponding to the partition function in the $X$=$A$,$D$,$E$ series
(so that $X$=$A$ for $p$=3,4, ~$X$=$A$ or $D$ for $p$$\geq$5, and
for $p$=11,12,17,18,29,30 also  $X$=$E$ is possible),
are given in~\rModInv.
Remarkably, we find the following ``completeness rules'' for arbitrary
$L$:
\eqn\comprule{\eqalign{
   Z\sL_{p(A)}(1) ~&=~{1\o 2}~{\rm Tr}~(I_{A_p})^{2L}
     ~~~~~~~~~~~~~~~~~~~~~~~~{\rm for}~~p=3,4,5,\ldots, \cr
   Z\sL_{p(D)}(1) ~&=~{1\o 2}~{\rm Tr}~(I_{D_{(p+3)/2}})^{2L}
     ~~~~~~~~~~~~~~~~~{\rm for}~~p=5,7,9,\ldots, \cr
   Z\sL_{p(E)}(1) ~&=~{1\o 2}~{\rm Tr}~(I_{E_n})^{2L}
     ~~~~~~~{\rm with}~~n=6,7,8~~{\rm for}~~p=11,17,29,~{\rm resp.,}\cr}}
where $I_X$ is the incidence matrix of the Dynkin diagram of the
Lie algebra $X$, whose rank is such that its Coxeter number
is $p$+1. (The traces can be evaluated explicitly using the known
eigenvalues of the $I_X$, which are given by ~$2\cos {\pi m\o g}$~ where
the $m$ and $g$ are the Coxeter exponents and Coxeter number of $X$,
respectively; using this fact, the particularly simple rhs's of
\Zthqone, \Zfiqone, and \Ztetqone\ can easily be seen to follow from
\comprule.)
The general $A_p$ case of these identities   can be proved
using \CpLqone, assuming \conj\ holds. For the
$D$- and $E$-cases we conjecture them based on their validity
for many small values of $p$ and $L$, which we have verified.
[One can also finitize the fermionic semi-modular-invariant
partition functions of~\rher, using a generalization of \finptfc.
We denote the resulting functions by $Z_{p({\rm f})}\sL(q)$, where
the $N_{r,s;\bar{r},\bar{s}}^{p({\rm f})}$ are read off from
the `fermionic $D$-series' expressions in eq.~(5.2) of~\rher,
so that ~$Z_{3({\rm f})}\sL(q)=Z_{\rm Maj}\sL(q)$~ of \finZth\
and \finZthpr. Results for small $p$ and $L$ lead us to conjecture
that ~$Z_{p({\rm f})}\sL(1)=Z_{p(D)}\sL(1)$~ for all odd $p\geq 5$,
generalizing \Zthqone.]

Note that the rhs's in \comprule\ are precisely half the number
of allowed height configurations along a row of $2L$ sites
(with periodic boundary conditions $h_{2L+1}=h_1$) of the RSOS
models~\rPasq~based on the Dynkin diagrams of
$A_p$, $D_{(p+3)/2}$ and $E_n$.
Since the lhs's are defined in terms of conformal field theory
characters, which have been ``finitized'' via a procedure
inspired by a quasi-particle description of the spectrum
of spin chains, the identities \comprule\ reflect an
intriguing connection between all these frameworks.

\medskip

Finally, let us return to the initial motivation for
the present work, namely proving the
conjectured~\rDKMM\rKKMMii~fermionic representations
\chieqS\ (and their extensions \chieqSt) for the Virasoro characters
of ${\cal M}(p,p+1)$.
As demonstrated, we have achieved this goal for $p$=4 using
the finitization procedure. However, for bigger $p$ this procedure
seems to be increasingly cumbersome and we have not been able yet
to fully carry it out and obtain a proof.

As a possible hint for an alternative way of attacking the problem,
which does not require finitization of the $q$-series involved,
we would like to draw the reader's attention to an interesting
feature of the fermionic $q$-series on the rhs's of \chieqS\ and
\chieqSt: in all cases the vector $\bA$, specifying the linear
shift of the quadratic form in the exponent of $q$, is either
zero or a unit vector.
(In fact, the extension \chieqSt\ of the conjecture \chieqS\
of~\rKKMMii~was arrived at by searching for possible right
choices of $\bQ,\bu$, and $\bA$, allowing for other types of
vectors; our -- clearly not exhaustive -- search did not yield
any viable fermionic sum representations where $\bA$ is not a unit
vector.)
This observation suggests that we consider the following formal series in
$n$ variables $z_a$ in addition to $q$,
\eqn\Snz{ S_n^{\bQ} (\bu|q,\bz ) =
 \sum_{ { {\bf m}\in 2\ZZ^n+{\bf Q} \atop m_1\geq 0}  }
  q^{{1\o 4}{\bf m} C_n {\bf m}^t}
   \left(\prod_{a=1}^n z_a^{-{1\o 2}m_a} \right)
  {1\o (q)_{m_1}}~\prod_{a=2}^n ~
  { {1\o 2}({\bf m}I_n+{\bf u})_a \atopwithdelims[] m_a}_q~~,}
which reduces to $S_n{\bQ \atd \bA}(\bu|q)$ of \Sn\ when
specializing to $z_a$=$q^{A_a}$~ for ~$a$=1, $\ldots$, $n$.
When inserted into \chieqS\ and
\chieqSt, this means that all the $z_a$ are in fact evaluated at
1, except possibly for one which is evaluated at $q$,
since $\bA_s=\be_{n+2-s}$. The series $S_n^{\bQ}(\bu|q,\bz )$
are then generalizations of
\eqn\rsrz{ S(q,z) ~=~ \sum_{m=0}^\infty {q^{m^2} z^m\o (q)_m}~~,}
giving the two fermionic sums with $a$=0 and 1
in \rsr\ when evaluated at $z$=1 and $z$=$q$, respectively.
This function, which satisfies a linear
second order $q$-difference equation,
plays an important role in the
proofs~\rRR\rSchur~of the Rogers-Ramanujan-Schur identities \rsr.

\vskip 18mm

{\vbox{\centerline{\bf Acknowledgements}}}
\nobreak\medskip \no
I would like to thank G.~Albertini, R.~Kedem, T.R.~Klassen, and
B.M.~McCoy for useful discussions.
This work was supported by the NSF, grant 91-08054.

\vfill\eject

\listrefs

\vfill\eject

\bye\end